\documentclass[a4paper]{jpconf}
\usepackage{amsmath,amssymb,amsthm}
\usepackage[dvips]{graphicx}

\newtheorem{theorem}{Theorem}

\theoremstyle{definition}

\theoremstyle{remark}

\newcommand{\beq}{\begin{eqnarray}}
\newcommand{\eeq}{\end{eqnarray}}
\newcommand{\beqnn}{\begin{eqnarray*}}
\newcommand{\eeqnn}{\end{eqnarray*}}
\newcommand{\rd}{\partial}

\newcommand{\tp}[1]{\:{}^{\mathrm{t}}#1}

\newcommand{\ZZ}{\mathbf{Z}}
\newcommand{\bst}{\boldsymbol{t}}
\newcommand{\bsx}{\boldsymbol{x}}
\newcommand{\bszero}{\boldsymbol{0}}

\newcommand{\calP}{\mathcal{P}}
\newcommand{\bstbar}{\bar{\bst}}
\newcommand{\tbar}{\bar{t}}
\newcommand{\Wbar}{\bar{W}}
\newcommand{\Lbar}{\bar{L}}
\begin{document}

\title{Modified melting crystal model and 
Ablowitz-Ladik hierarchy}
\author{Kanehisa Takasaki}
\address{Human and Environmental Studies, 
Kyoto University, Kyoto 606-8501, Japan}
\ead{takasaki@math.h.kyoto-u.ac.jp}

\begin{abstract}
This is a review of recent results on the integrable structure 
of the ordinary and modified melting crystal models.  
When deformed by special external potentials, 
the partition function of the ordinary melting crystal model 
is known to become essentially a tau function of the 1D Toda hierarchy. 
In the same sense, the modified model turns out to be related 
to the Ablowitz-Ladik hierarchy.  These facts are explained 
with the aid of a free fermion system, fermionic expressions 
of the partition functions, algebraic relations among 
fermion bilinears and vertex operators, 
and infinite matrix representations of those operators.  
\end{abstract}

\section{Introduction}
\medskip

Recently, we extended our previous work \cite{NT07,NT08} 
on the integrable structure of the melting crystal model 
--- a model of {\it random 3D Young diagrams} --- 
to a modified model \cite{Takasaki12,Takasaki13}.  
In the context of string theory \cite{Marino-book}, 
the modified model is related to topological string theory 
on the so called {\it resolved conifold}.  
Our study was motivated by Brini's conjecture \cite{Brini10} 
that all genus Gromov-Witten invariants of this 
non-compact complex Calabi-Yau 3-fold can be captured 
by the Ablowitz-Ladik hierarchy \cite{AL75}.  
Brini confirmed his conjecture to the first order 
(i.e., genus $\le 1$) of genus expansion. 
Proving the conjecture in the full genus case 
remained open.  We believe that our results show 
an affirmative answer (or, at least, a decisive piece 
of evidence) to this question.  

In our previous work \cite{NT07,NT08}, the partition function 
of the melting crystal model is shown to be related to a tau function 
of the 1D Toda hierarchy.  This fact is proven with the aid 
of a free fermion system, a fermionic realization 
of the {\it quantum torus algebra}, and algebraic relations 
called {\it shift symmetries} in this algebra.  
We could use these tools for the modified model as well 
to show that the partition function is related 
to a tau function of the 2D Toda hierarchy \cite{Takasaki12}.  
Although this was an important step, a new technique was 
wanted to reach the goal. 

A final clue was found in the work of Brini et al. \cite{BCR11}. 
They presented a characterization of the Ablowitz-Ladik hierarchy 
(or, rather, the relativistic Toda hierarchy \footnote{
These two integrable hierarchies are known to be equivalent 
\cite{KMZ96,Suris97}, but it is the relativistic Toda hierarchy 
that can be embedded into the 2D Toda hierarchy in a natural manner.})
in the Lax formalism of the 2D Toda hierarchy.  
According to the characterization by Brini et al., 
the Ablowitz-Ladik hierarchy amounts to the case 
where the Lax operators of the 2D Toda hierarchy 
have a special {\it quotient} (or {\it rational\/}) form.   
We examined the Lax operators of the aforementioned tau function, 
and confirmed that they do have the special form \cite{Takasaki13}.

\section{Melting crystal model}
\medskip

\subsection{Crystal corner, 3D Young diagrams and plane partitions}
\medskip

\begin{figure}[b]
\begin{minipage}[b]{15pc}
\begin{center}
\includegraphics[scale=0.25]{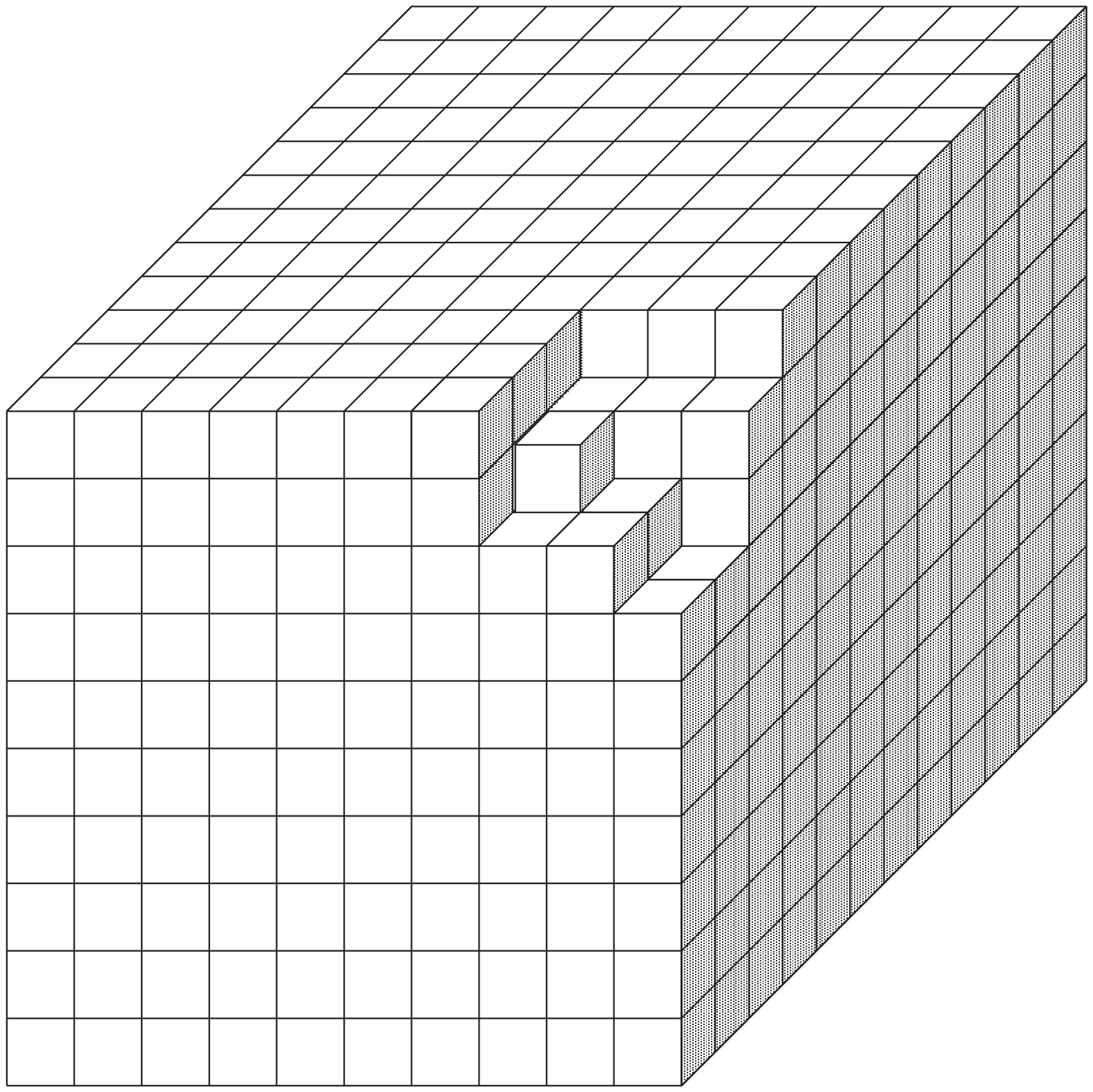}
\caption{\label{fig:crystal}Melting crystal corner}
\end{center}
\end{minipage}
\begin{minipage}[b]{15pc}
\begin{center}
\includegraphics[scale=0.8]{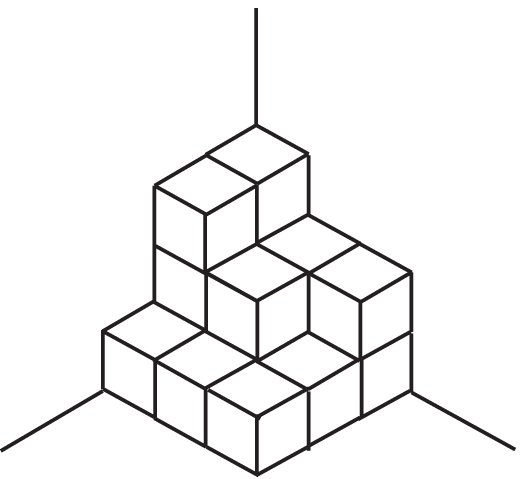}
\caption{\label{fig:3DYD}3D Young diagram}
\end{center}
\end{minipage}
\end{figure}

The melting crystal model is a statistical model 
of a crystal corner (Figure \ref{fig:crystal}). 
The crystal consists of unit cubes and occupies 
an octant of the $xyz$ space.  
The complement of the crystal in the octant 
is a {\it 3D Young diagram}, which conversely 
determines the shape of the crystal (Figure \ref{fig:3DYD}). 

Just as ordinary Young diagrams are identified 
with ordinary partitions 
\beqnn
  \lambda = (\lambda_i)_{i=1}^\infty, \quad 
  \lambda_1 \ge \lambda_2 \ge \ldots \ge 0, \quad
  |\lambda| = \sum_{i=1}^\infty\lambda_i < \infty, 
\eeqnn
of integers \cite{Macdonald-book}, 3D Young diagrams 
are in one-to-one correspondence with {\it plane partitions}.  
Plane partitions are 2D arrays of non-negative integers 
that are decreasing in two directions: 
\beqnn
\pi = (\pi_{ij})_{i,j=1}^\infty
= \left(\begin{array}{ccc}
  \pi_{11} & \pi_{12} & \cdots \\
  \pi_{21} & \pi_{22} & \cdots \\
  \vdots   & \vdots   & \ddots 
  \end{array}\right), 
  \quad 
  \begin{array}{ccc}
  \pi_{ij} &\ge& \pi_{i,j+1}\\
  \mbox{\rotatebox[origin=c]{-90}{$\ge$}} && \\
  \pi_{i+1,j} &&
  \end{array}, 
  \quad 
  |\pi| = \sum_{i,j=1}^\infty \pi_{ij} < \infty. 
\eeqnn
Such a plane partition determines a 3D Young diagram 
that consists of stacks of unit cubes of height $\pi_{ij}$
on the unit squares $[i-1,i]\times[j-1,j]$, 
$i,j = 1,2,\ldots$, of the $xy$ plane.  For example, 
the 3D Young diagram of Figure \ref{fig:3DYD} 
corresponds to 
\beqnn
  \pi = \left(\begin{array}{ccc}
        3 & 2 & 2 \\
        3 & 2 & 1 \\
        1 & 1 & 1 
        \end{array}\right). 
\eeqnn

The partition function of this model is the sum 
\beq
  Z = \sum_{\pi\in\calP\calP} q^{|\pi|}
\label{Z-def}
\eeq
of the Boltzmann weight $q^{|\pi|}$ ($0 < q < 1$)
over the set $\calP\calP$ of all plane partitions.

\subsection{Method of diagonal slicing} 

\begin{figure}[b]
\begin{minipage}[b]{15pc}
\begin{center}
\includegraphics[scale=0.8]{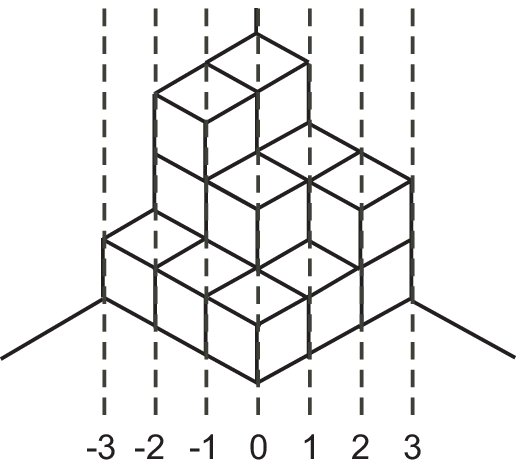}
\caption{\label{fig:slices}Diagonal slices}
\end{center}
\end{minipage}
\begin{minipage}[b]{15pc}
\begin{center}
\includegraphics[scale=0.6]{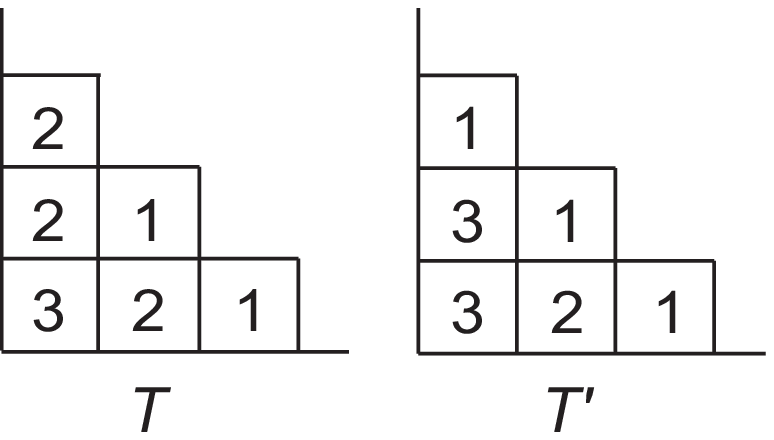}
\caption{\label{fig:tableaux}Semi-standard tableaux}
\end{center}
\end{minipage}
\end{figure}

The partition function (\ref{Z-def}) can be calculated 
by the method of {\it diagonal slicing} \cite{OR01}. 
Let us define the $m$-th diagonal slice $\pi(m)$, 
$m \in \ZZ$, of $\pi$ as 
\beqnn
  \pi(m) = 
  \begin{cases}
    (\pi_{i,i+m})_{i=1}^\infty &\mbox{if $m \ge 0$},\\
    (\pi_{j-m,j})_{j=1}^\infty &\mbox{if $ m < 0$}.  
  \end{cases}
\eeqnn
$\pi(m)$ represents the slice of the 3D Young diagram 
along the vertical plane $y = x + m$ (Figure \ref{fig:slices}). 
The two sets $\{\pi(-m)\}_{m=0}^\infty$ 
and $\{\pi(m)\}_{m=0}^\infty$ of these slices 
give two increasing sequences of Young diagrams 
that fill up the Young diagram of shape $\lambda = \pi(0)$. 

These increasing sequences of Young diagrams can be encoded 
to {\it semi-standard tableaux} $T,T'$ of shape $\lambda$ 
\cite{Macdonald-book}. 
These tableaux are obtained by entering the numbers $m+1$, 
$m = 0,1,2,\ldots$, to the boxes of the skew Young diagrams 
$\pi(-m)/\pi(-m-1)$ and $\pi(m)/\pi(m+1)$ (Figure \ref{fig:tableaux}). 
It is easy to see that the entries $T(i,j)$ and $T'(i,j)$ 
of $T,T'$ satisfy the conditions 
\beq
  \begin{array}{ccc}
  T(i,j) &\ge& T(i,j+1)\\
  \mbox{\rotatebox[origin=c]{-90}{$>$}} && \\
  T(i+1,j) &&
  \end{array},
  \quad
  \begin{array}{ccc}
  T'(i,j) &\ge& T'(i,j+1)\\
  \mbox{\rotatebox[origin=c]{-90}{$>$}} && \\
  T'(i+1,j) &&
  \end{array} 
\label{semi-standard}
\eeq
that characterize semi-standard tableaux \footnote{
This is different from the common definition 
of semi-standard tableaux \cite{Macdonald-book} 
in which all inequalities of (\ref{semi-standard}) 
are reversed.  Since Schur functions are symmetric 
functions, this difference does not affect 
the combinatorial definition (\ref{Schur-def}) 
of Schur functions.}.  
Thus any plane partition $\pi$ determines 
a triple $(\lambda,T,T')$ of a partition $\lambda$ 
and semi-standard tableaux $T,T'$ of shape $\lambda$.  
It is also easy to see that any such triple 
determine a plane partition.  The mapping 
$\pi \to (\lambda,T,T')$ thus turns out 
to be a bijection between $\calP\calP$ 
and the set of all triples. 

The sum in (\ref{Z-def}) can be thus converted to a sum 
over $(\lambda,T,T')$'s.  Moreover, the weight $q^{|\pi|}$ 
can be factorized as 
\beq
  q^{|\pi|} = q^Tq^{T'},\quad 
  q^T = \prod_{m=0}^\infty q^{(m+1/2)|\pi(-m)/\pi(-m-1)|},\quad 
  q^{T'} = \prod_{m=0}^\infty q^{(m+1/2)|\pi(m)/\pi(m+1)|}. 
\eeq
Because of this factorization, the sum in (\ref{Z-def}) 
can be separated to partial sums with respect to $T,T'$ 
and a sum with respect to $\lambda$ as 
\beqnn
  Z = \sum_{\lambda\in\calP}
      \left(\sum_{T:\;\mathrm{shape}\;\lambda}q^T\right) 
      \left(\sum_{T':\;\,\mathrm{shape}\;\lambda}q^{T'}\right), 
\eeqnn
where $\calP$ denotes the set of all partitions.  
According to the combinatorial definition 
\beq
  s_\lambda(\bsx) = \sum_{T:\;\mathrm{shape}\;\lambda}\bsx^T,\quad 
  \bsx^T = \prod_{(i,j)}x_{T(i,j)}, 
\label{Schur-def}
\eeq
of Schur functions of $\bsx = (x_1,x_2,\ldots)$ \cite{Macdonald-book}, 
the partial sums over $T,T'$ become a special value 
of $s_\lambda(\bsx)$: 
\beq
    \sum_{T:\;\mathrm{shape}\;\lambda}q^T
  = \sum_{T':\;\mathrm{shape}\;\lambda}q^{T'}
  = s_\lambda(q^{-\rho}), \quad 
  q^{-\rho} = \left(q^{1/2},q^{3/2},\ldots,q^{k+1/2},\ldots\right). 
\eeq

The partition function can be thus reduced to a sum over $\calP$: 
\beq
  Z = \sum_{\lambda\in\calP}s_\lambda(q^{-\rho})^2. 
\label{Z=sum_lambda}
\eeq
One can now use the Cauchy identity \cite{Macdonald-book}
\beq
  \sum_{\lambda\in\calP}
  s_\lambda(x_1,x_2,\ldots)s_\lambda(y_1,y_2,\ldots)
  = \prod_{i,j=1}^\infty (1 - x_iy_j)^{-1} 
\eeq
to calculate (\ref{Z=sum_lambda}) explicitly: 
\beq
  Z = \sum_{\lambda\in\calP}s_\lambda(q^{-\rho})^2 
    = \prod_{i,j=1}^\infty(1 - q^{i+j-1})^{-1}
    = \prod_{n=1}^\infty (1-q^n)^{-n}. 
\eeq
This function is known as the {\it MacMahon function\/}.

\subsection{Deformations by external potentials}
\medskip

A simplest deformation of this model is obtained 
by inserting the extra weight $Q^{|\pi(0)|}$, 
where $Q$ is a new positive constant. 
The foregoing partition function (\ref{Z-def}) 
is thereby deformed as 
\beq
  Z = \sum_{\pi\in\calP\calP}q^{|\pi|}Q^{|\pi(0)|}. 
\eeq
This is a deformation induced by the external potential 
$|\pi(0)|\log Q$.  By the mapping $\pi \mapsto (\lambda,T,T')$, 
this sum, too, can be reduced to a sum of the form 
\beq
  Z = \sum_{\lambda\in\calP}s_\lambda(q^{-\rho})^2Q^{|\lambda|}, 
\label{Z(Q)=sum_lambda}
\eeq
and eventually boils down to a generalization 
of the MacMahon function: 
\beq
  Z = \prod_{i,j=1}^\infty(1 - Qq^{i+j-1})^{-1} 
    = \sum_{n=1}^\infty (1-Qq^n)^{-n}. 
\eeq

An integrable hierarchy emerges in a deformed 
partition function $Z(s,\bst)$ that depends 
on a discrete variable $s \in \ZZ$ 
and a set of continuous variables $\bst = (t_1,t_2,\ldots)$.  
The deformations are induced by the external potential 
\beq
  \Phi(\lambda,s,\bst) 
  = \sum_{k=1}^\infty t_k\Phi_k(\lambda,s), \quad 
  \Phi_k(\lambda,s) 
  = \sum_{i=1}^\infty(q^{k(\lambda_i+s-i+1)} - q^{k(s-i+1)}) 
   + \frac{1-q^{ks}}{1-q^k}q^k. 
\eeq
Note that the sum in the definition of $\Phi_k(\lambda,s)$ 
is actually a finite sum.  This expression is an analytic 
continuation of the expression 
\beq
  \Phi_k(\lambda,s) 
  = \sum_{i=1}^\infty q^{k(\lambda_i+s-i+1)} - \sum_{i=1}^\infty q^{k(-i+1)} 
\eeq
in the domain $|q| > 1$ of the $q$-plane.  
The term $Q^{|\lambda|}$ in (\ref{Z(Q)=sum_lambda}), too, has to be 
replaced by the $s$-dependent form $Q^{|\lambda|+s(s+1)/2}$. 
With the aid of 2D complex free fermion system 
(which will be briefly reviewed in the next section), 
we obtained the following result  \cite{NT07,NT08}: 

\begin{theorem}
The deformed partition function 
\beq
  Z(s,\bst) 
  = \sum_{\lambda\in\calP}
    s_\lambda(q^{-\rho})^2 Q^{|\lambda|+s(s+1)/2}e^{\Phi(\lambda,s,\bst)} 
\label{Z(s,t)-def}
\eeq
is related to a special tau function $\tau(s,\bst)$ 
of the 1D Toda hierarchy as 
\beq
  Z(s,\bst) 
  = \exp\left(\sum_{k=1}^\infty\frac{t_kq^k}{1-q^k}\right)
    q^{-s(s+1)(2s+1)/6}\tau(s,\iota(\bst)), 
\label{Z=tau}
\eeq
where $\iota(\bst)$ denotes the alternating inversion 
\beqnn
  \iota(\bst) = (-t_1, t_2, -t_3, \ldots, (-1)^kt_k,\ldots) 
\eeqnn
of $\bst = (t_1,t_2,\ldots)$.  
\end{theorem}
\medskip

\section{Modified melting crystal model}
\medskip

\subsection{Partition function}
\medskip

The modified model is obtained by replacing 
\beqnn
  s_\lambda(q^{-\rho})^2 \;\to\; 
  s_\lambda(q^{-\rho})s_{\tp{\lambda}}(q^{-\rho}),
\eeqnn
where $\tp{\lambda}$ denotes the conjugate partition
that represents the transpose of the associated Young diagram. 
The $Q$-dependent partition function reads 
\beq
  Z' = \sum_{\lambda\in\calP}
       s_\lambda(q^{-\rho})s_{\tp{\lambda}}(q^{-\rho})Q^{|\lambda|}. 
\label{Z'(Q)-def}
\eeq
This is no longer a statistical model of 3D Young diagrams.  
$s_\lambda(q^{-\rho})$ and $s_{\tp{\lambda}}(q^{-\rho})$ 
are partial sums of contributions from ``half'' pieces 
of 3D Young diagrams, but these pieces cannot be glued 
along the plane $x = y$. By the Cauchy identity
\beq
  \sum_{\lambda\in\calP}
  s_\lambda(x_1,x_2,\ldots)s_{\tp{\lambda}}(y_1,y_2,\ldots) 
  = \prod_{i,j=1}^\infty (1 + x_iy_j)
\eeq
of another type \cite{Macdonald-book}, 
one can convert (\ref{Z'(Q)-def}) to an infinite product 
of the form 
\beq
  Z' = \sum_{i,j=1}^\infty (1 + Qq^{i+j-1}) 
     = \prod_{n=1}^\infty (1 + Qq^n)^n. 
\eeq

We deform (\ref{Z'(Q)-def}) by the external potential 
\beqnn
  \Phi(\lambda,s,\bst,\bstbar) 
  = \sum_{k=1}^\infty t_k\Phi_k(\lambda,s) 
    + \sum_{k=1}^\infty\tbar_k\Phi_{-k}(\lambda,s) 
\eeqnn
that depends on a discrete variable $s$ and 
two sets of continuous variables 
$\bst = (t_1,t_2,\ldots)$ and 
$\bstbar = (\tbar_1,\tbar_2,\ldots)$.  
The deformed partition function reads 
\beq
  Z'(s,\bst,\bstbar) 
  = \sum_{\lambda\in\calP}
    s_\lambda(q^{-\rho})s_{\tp{\lambda}}(q^{-\rho})
    Q^{|\lambda|+s(s+1)/2} e^{\Phi(\lambda,s,\bst,\bstbar)}. 
\label{Z'(s,t,tbar)-def}
\eeq
To identify an integrable hierarchy hidden therein, 
we translate the partition function 
to the language of the same 2D complex free fermion system 
as used in our previous work \cite{NT07,NT08}.

\subsection{Perspectives from fermions}
\medskip

Let $\psi_n$ and $\psi^*_n$, $n \in \ZZ$, be the Fourier modes 
of 2D complex free fermion fields \cite{MJD-book}
\beqnn
  \psi(z) = \sum_{n\in\ZZ}\psi_nz^{-n-1},\quad 
  \psi^*(z) = \sum_{n\in\ZZ}\psi^*_nz^{-n}. 
\eeqnn
They satisfy the anti-commutation relations 
\beqnn
   \psi_m\psi^*_n + \psi^*_n\psi_n = \delta_{m+n,0},\quad 
   \psi_m\psi_n + \psi_n\psi_m = \psi^*_m\psi^*_n + \psi^*_n\psi^*_m = 0. 
\eeqnn
The associated Fock and dual Fock space can be decomposed 
to charge-$s$ ($s \in \ZZ$) sectors.  
Let $\langle s|$ and $|s\rangle$ denote 
the normalized ground states in the charge-$s$ sector: 
\beqnn
  \langle s| = \langle-\infty|\cdots\psi^*_{s-1}\psi^*_s,\quad 
  |s\rangle = \psi_{-s}\psi_{-s+1}\cdots|-\infty\rangle
\eeqnn
Excited states are labelled by partitions $\lambda \in \calP$ 
as $\langle \lambda,s|$ and $|\lambda,s\rangle$:
\beqnn
  \langle\lambda,s| 
  = \langle-\infty|\cdots\psi^*_{\lambda_2+s-1}\psi^*_{\lambda_1+s},\quad 
  |\lambda,s\rangle 
  = \psi_{-\lambda_1-s}\psi_{-\lambda_2-s+1}\cdots|-\infty\rangle.
\eeqnn

Let us introduce the fermion bilinears 
\beqnn
\begin{gathered}
  L_0 = \sum_{n\in\ZZ}n{:}\psi_{-n}\psi^*_n{:},\quad
  W_0 = \sum_{n\in\ZZ}n^2{:}\psi_{-n}\psi^*_n{:},\\
  H_k = \sum_{n\in\ZZ}q^{kn}{:}\psi_{-n}\psi^*_n{:},\quad
  J_k = \sum_{n\in\ZZ}{:}\psi_{-n}\psi^*_{n+k}{:}, 
\end{gathered}
\eeqnn
the vertex operators \cite{OR01,BY08}
\beqnn
  \Gamma_{\pm}(z) 
    = \exp\left(\sum_{k=1}^\infty\frac{z^k}{k}J_{\pm k}\right),\quad
  \Gamma'_{\pm}(z) 
    = \exp\left(-\sum_{k=1}^\infty\frac{(-z)^k}{k}J_{\pm k}\right), 
\eeqnn
and the infinite products 
\beqnn
  \Gamma_{\pm}(q^{-\rho}) = \prod_{i=1}^\infty\Gamma_{\pm}(q^{i-1/2}),\quad 
  \Gamma'_{\pm}(q^{-\rho}) = \prod_{i=1}^\infty\Gamma'_{\pm}(q^{i-1/2}) 
\eeqnn
of the vertex operators specialized to $z = q^{i-1/2}$, $i = 1,2,\ldots$.  
Building blocks of (\ref{Z'(s,t,tbar)-def}) can be expressed 
with these operators as 
\beq
\begin{gathered}
  s_\lambda(q^{-\rho}) 
    = \langle s|\Gamma_{+}(q^{-\rho})|\lambda,s\rangle,\quad
  s_{\tp{\lambda}}(q^{-\rho}) 
    = \langle\lambda,s|\Gamma'(q^{-\rho})|s\rangle,\\
  Q^{|\lambda|+s(s+1)/2} = \langle\lambda,s|Q^{L_0}|\lambda,s\rangle,\quad
  \Phi_k(\lambda,s) = \langle\lambda,s|H_k|\lambda,s\rangle. 
\end{gathered}
\eeq
Note that $L_0$ and $H_k$ are diagonal with respect to 
the orthonormal bases $\langle\lambda,s|$ and $|\lambda,s\rangle$ 
($\lambda\in\calP$, $s\in\ZZ$) of the Fock and dual Fock spaces. 
Thus (\ref{Z'(s,t,tbar)-def}) turns out to be expressed as 
\beq
  Z'(s,\bst,\bar{\bst})  
  = \langle s|\Gamma_{+}(q^{-\rho})Q^{L_0}e^{H(\bst,\bar{\bst})} 
    \Gamma'_{-}(q^{-\rho})|s\rangle, 
\eeq
where 
\beqnn
  H(\bst,\bar{\bst}) 
  = \sum_{k=1}^\infty t_kH_k + \sum_{k=1}^\infty\bar{t}_kH_{-k}. 
\eeqnn

For comparison, let us recall that the deformed partition function 
(\ref{Z(s,t)-def}) of the previous model has a similar 
fermionic expression 
\beq
  Z(s,\bst) 
  = \langle s|\Gamma_{+}(q^{-\rho})Q^{L_0}e^{H(\bst)} 
    \Gamma_{-}(q^{-\rho})|s\rangle, 
  \quad 
  H(\bst) = \sum_{k=1}^\infty t_kH_k. 
\eeq
In our previous work \cite{NT07,NT08}, we derived (\ref{Z=tau}) 
from this fermionic expression of $Z(s,\bst)$ 
by manipulation of operators on the fermionic Fock space.  
The tau function $\tau(s,\bst)$ is defined by the fermionic formula 
\beq
  \tau(s,\bst) 
  = \langle s|\exp\left(\sum_{k=1}^\infty t_kJ_k\right)g|s \rangle 
  = \langle s|g\exp\left(\sum_{k=1}^\infty t_kJ_{-k}\right)|s\rangle, 
\label{fermionic-1Dtau}
\eeq
where $g$ is an operator of the form 
\beq
  g = q^{W_0/2}\Gamma_{-}(q^{-\rho})\Gamma_{+}(q^{-\rho})Q^{L_0}
      \Gamma_{-}(q^{-\rho})\Gamma_{+}(q^{-\rho})q^{W_0/2}.
\eeq
The existence of two apparently different expressions 
in (\ref{fermionic-1Dtau}) is a consequence 
of the algebraic relations 
\beq
  J_kg = gJ_{-k}, \quad k = 1,2,\ldots. 
\eeq

Extending this work, we have been able to show 
the following result \cite{Takasaki12}. 

\begin{theorem}
The partition function $Z'(s,\bst,\bstbar)$ 
of the modified model is related to 
a tau function $\tau'(s,\bst,\bar{\bst})$ 
of the 2D Toda hierarchy as 
\beq
  Z'(s,\bst,\bar{\bst}) 
  = \exp\left(\sum_{k=1}^\infty\frac{q^kt_k - \bar{t}_k}{1-q^k}\right)
    \tau'(s,\iota(\bst),- \bar{\bst}). 
\label{Z'=tau'}
\eeq
$\tau'(s,\bst,\bar{\bst})$ is defined by the fermionic formula 
\beq
  \tau'(s,\bst,\bar{\bst}) 
  = \langle s|\exp\left(\sum_{k=1}^\infty t_kJ_k\right)
    g'\exp\left(- \sum_{k=1}^\infty\bar{t}_kJ_{-k}\right)|s\rangle, 
\label{fermionic-2Dtau}
\eeq
where $g'$ is an operator of the form 
\beq
  g' = q^{W_0/2}\Gamma_{-}(q^{-\rho})\Gamma_{+}(q^{-\rho})Q^{L_0}
       \Gamma'_{-}(q^{-\rho})\Gamma'_{+}(q^{-\rho})q^{-W_0/2}.  
\eeq
\end{theorem}

Derivation of this result is parallel to the case of 
$Z(s,\bst)$ \cite{NT07,NT08}.  ``Shift symmetries'' 
in the quantum torus algebra of fermion bilinears 
imply the following algebraic relations ($k > 0$): 
\beq
\begin{gathered}
  \Gamma_{+}(q^{-\rho})H_k\Gamma_{+}(q^{-\rho})^{-1}
  = (-1)^k\Gamma_{-}(q^{-\rho})^{-1}
    q^{-W_0/2}J_kq^{W_0/2}\Gamma_{-}(q^{-\rho}) 
    + \frac{q^k}{1-q^k},\\
  \Gamma'_{-}(q^{-\rho})^{-1}H_{-k}\Gamma'_{-}(q^{-\rho}) 
  = \Gamma'_{+}(q^{-\rho})q^{-W_0/2}J_{-k}
    q^{W_0/2}\Gamma'_{+}(q^{-\rho})^{-1}
    - \frac{1}{1-q^k}. 
\end{gathered}
\eeq
With the aid of these algebraic relations, one can derive 
(\ref{Z=tau}) and (\ref{Z'=tau'}) by rather straightforward 
calculations. 

This is not the end of the story.  To uncover 
hidden properties of $\tau'(s,\bst,\bstbar)$, 
we now translate the fermionic machinery 
to the language of infinite matrices.

\subsection{From fermions to infinite matrices}
\medskip

It is well known \cite{MJD-book} that fermion bilinears 
are in one-to-one correspondence with $\ZZ\times\ZZ$ matrices as 
\beqnn
  \hat{X} =   \sum_{i,j\in\ZZ}x_{ij}{:}\psi_{-i}\psi^*_j{:} 
  \;\longleftrightarrow\;
  X = \sum_{i,j\in\ZZ}x_{ij}E_{ij}. 
\eeqnn
Apart from c-number corrections, they obey 
the same commutation relations, namely, 
\beq
  [\hat{X},\hat{Y}] = \widehat{[X,Y]} + c(X,Y), 
\eeq
where $c(X,Y)$ is a c-number term.  

Actually, the set of infinite matrices is equipped 
with multiplication as well as Lie brackets.  
This provides us with more freedom. 
For example, matrix representation of 
our fundamental fermion bilinears $L_0,M_0,H_k,J_k$ 
(let us use the same notations as fermion bilinears) 
can be written as 
\beq
  L_0 = \Delta,\quad 
  W_0 = \Delta^2,\quad 
  H_k = q^{k\Delta},\quad 
  J_k = \Lambda^k, 
\eeq
where 
\beqnn
  \Delta = \sum_{n\in\ZZ}nE_{nn},\quad 
  \Lambda = \sum_{n\in\ZZ}E_{n,n+1},\quad 
  E_{mn} = (\delta_{im}\delta_{jn})_{i,j\in\ZZ}. 
\eeqnn
Matrix representation of the vertex operators 
reveals an even more significant feature. 
Since 
\beq
  \sum_{k=1}^\infty\frac{z^k}{k}\Lambda^{\pm 1} 
    = - \log(1 - z\Lambda^{\pm}),\quad 
  \sum_{k=1}^\infty\frac{(-z)^k}{k}\Lambda^{\pm k} 
    = - \log(1 + z\Lambda^{\pm}), 
\eeq
$\Gamma_{\pm}(z)$ and $\Gamma_{\pm}'(z)$ can be expressed as 
\beq
  \Gamma_{\pm}(z) = (1 - z\Lambda^{\pm 1})^{-1},\quad 
  \Gamma'_{\pm}(z) = 1 + z\Lambda^{\pm 1}, 
\eeq
hence 
\beq
  \Gamma_{\pm}(q^{-\rho}) 
  = \prod_{i=1}^\infty (1 - q^{i-1/2}\Lambda^{\pm 1})^{-1},\quad
  \Gamma'_{\pm}(q^{-\rho}) 
  = \prod_{i=1}^\infty (1 + q^{i-1/2}\Lambda^{\pm 1}). 
\eeq
The latter may be thought of as matrix-valued 
{\it quantum dilogarithm} \cite{FV93,FK93}.  
Moreover, these vertex operators show up 
in $g$ and $g'$ in such a form 
as $\Gamma_{-}(q^{-\rho})\Gamma_{+}(q^{-\rho})$ 
and $\Gamma_{-}'(q^{-\rho})\Gamma_{+}'(q^{-\rho})$.  
As one can see from Jacobi's triple product formula 
\beq
  \vartheta(z) 
  = \prod_{n=1}^\infty(1-q^n)
    \prod_{n=1}^\infty(1+q^{n-1/2}z)
    \prod_{n=1}^\infty(1+q^{n-1/2}z^{-1}), 
\eeq
this indicates a possible relation with the theta function. 
This issue deserves to be studied in more detail \footnote{
We learned from John Harnad that the same theta function 
emerges in his calculations on a class of tau functions. 
Those tau functions were studied by Sasha Orlov \cite{Orlov03}
in a different context.}.

\subsection{Perspectives from infinite matrices}
\medskip

The set of $\ZZ\times\ZZ$ matrices is the place 
where the Lax formalism of the 2D Toda hierarchy 
can be reformulated \cite{UT84}.  
Note that $\Delta$ and $\Lambda$ amount to 
the scalar operator $s$ and the shift operator $e^{\rd_s}$ 
that are fundamental in the Lax formalism 
based on difference operators \cite{TT95}.  
In this setting, one can consider a {\it factorization problem} 
that captures all solutions of the 2D Toda hierarchy 
\cite{Takasaki84}: 

\paragraph{Factorization problem} 
Given a constant invertible $\ZZ\times\ZZ$ matrix $U$, 
find two $\ZZ\times\ZZ$ matrices $W = W(\bst,\bstbar)$ 
and $\Wbar = \Wbar(\bst,\bstbar)$ that satisfy 
the following conditions: 
\begin{itemize}
\item $W$ is lower triangular, and all diagonal elements 
are equal to $1$. 
\item $\Wbar$ is upper triangular, and all diagonal elements 
are nonzero. 
\item They satisfy the factorization relation
\beq
  \exp\left(\sum_{k=1}^\infty t_k\Lambda^k\right)
    U \exp\left(- \sum_{k=1}^\infty\bar{t}_k\Lambda^{-k}\right) 
  = W^{-1}\Wbar.
\label{factorization}
\eeq
\end{itemize}

If one can find such a pair $W,\Wbar$, the Lax matrices 
\beq
  L = W\Lambda W^{-1}, \quad 
  \bar{L} = \Wbar\Lambda\Wbar^{-1}
\label{LLbar}
\eeq
satisfy the Lax equations (hence become a solution) 
of the 2D Toda hierarchy. 

To derive a solution that corresponds to $\tau'(s,\bst,\bstbar)$, 
we choose the matrix $U$ to be matrix representation of $g'$: 
\beq
  U = q^{\Delta^2/2}\Gamma_{-}(q^{-\rho})\Gamma_{+}(q^{-\rho})Q^{\Delta}
       \Gamma'_{-}(q^{-\rho})\Gamma'_{+}(q^{-\rho})q^{-\Delta^2/2}. 
\eeq
Remarkably, this matrix can be factorized as follows: 
\beq
  U = W^{-1}\bar{W},\quad 
  \begin{array}{l}
  W = q^{\Delta^2/2}\Gamma'_{-}(Qq^{-\rho})^{-1}
      \Gamma_{-}(q^{-\rho})^{-1}q^{-\Delta^2/2},\\
  \bar{W} = q^{\Delta^2/2}Q^\Delta\Gamma_{+}(Qq^{-\rho})
            \Gamma'_{+}(q^{-\rho})q^{-\Delta^2/2}. 
  \end{array}
\eeq
This means that the factors $W,\Wbar$ give a solution 
of the factorization problem (\ref{factorization}) 
at the ``initial time'' $\bst = \bstbar = \bszero$. 
This allows us to find an explicit form of 
the associated Lax matrices (\ref{LLbar}) as well: 
\beq
\begin{gathered}
  L = (\Lambda - q^\Delta)(1 + Qq^{\Delta-1}\Lambda^{-1})^{-1},\\
  \bar{L}^{-1} = (1 + Qq^{\Delta-1}\Lambda^{-1})(q^\Delta - \Lambda)^{-1}. 
\end{gathered}
\label{initial-LLbar}
\eeq
Thus the ``initial values'' of the Lax matrices 
at $\bst = \bstbar = \bszero$ take a very special form 
(so to speak, ``quotients'' of first order difference operators).  

As pointed out by Brini et al. \cite{BCR11}, 
the ``quotient'' ansatz 
\beq
\begin{gathered}
  L = BC^{-1}, \quad \Lbar^{-1} = C(-B)^{-1},\\
  B = e^{\rd_s} - b,\quad C = I - ce^{-\rd_s}, 
\end{gathered}
\eeq
of the Lax operators, where $b$ and $c$ are scalar-valued 
functions of $s$, $\bst$ and $\bstbar$, 
is preserved under time evolutions 
of the 2D Toda hierarchy. In other words, 
this is a reduction condition, and the reduced system 
is the Ablowitz-Ladik hierarchy \cite{AL75} 
(or the relativistic Toda hierarchy \cite{KMZ96,Suris97}). 
(\ref{initial-LLbar}) amount to Lax operators of this type.  
We are thus led to the following conclusion: 

\begin{theorem}
$\tau'(s,\bst,\bstbar)$ is the tau function 
of a solution of the Ablowitz-Ladik hierarchy 
embedded in the 2D Toda hierarchy.  
\end{theorem}

\section*{Acknowledgements}
\smallskip

We are very grateful to John Harnad for fruitful discussion. 
The remark on the theta function in Section 3.3 
is inspired by his comments. 
This work is partly supported by JSPS Grants-in-Aid 
for Scientific Research No. 24540223 and No. 25400111 
from the Japan Society for the Promotion of Science.

\end{document}